\documentclass[pra,showpacs,nofootinbib]{revtex4}
\usepackage{hyperref}
\usepackage{amsmath}
\usepackage{amssymb}
\usepackage{amsthm}
\usepackage{graphics}
\usepackage{graphicx}
\usepackage{color}
\usepackage[usenames,dvipsnames,svgnames,table]{xcolor}

\long\def\ca#1\cb{}

\def\bra#1{\langle#1|}
\def\dyad#1#2{|#1\rangle\langle#2|}

\def\ket#1{|#1\rangle }

\def\Tr#1{\textrm{Tr}(#1)}

\def\HC{{\cal H}}

\def\JC{{\cal J}}
\def\KC{{\cal K}}

\def\SC{{\cal S}}

\newtheorem{thm1}{Theorem}
\newtheorem{thm2}[thm1]{Theorem}

\newtheorem{lem1}[thm1]{Lemma}

\newtheorem{lem3}[thm1]{Lemma}

\newtheorem{def1}[thm1]{Definition}

\begin{document}
\title{Necessary condition for local quantum operations and classical communication with extensive violation by separable operations}
\author{Scott M. Cohen}
\email{cohensm52@gmail.com}
\affiliation{Department of Physics, Portland State University, Portland OR 97201}

\begin{abstract}
We give a conceptually simple necessary condition such that a separable quantum operation can be implemented by local operations on subsystems and classical communication between parties (LOCC), a condition which follows from a novel approach to understanding LOCC. This necessary condition holds for any number of parties and any finite number of rounds of communication and as such, also provides a completely general sufficient condition that a given separable operation cannot be exactly implemented by LOCC. Furthermore, it demonstrates an extremely strong difference between separable operations and LOCC, in that there exist examples of the former for which the condition is extensively violated. More precisely, the violation by separable operations of our necessary condition for LOCC grows without limit as the number of parties increases.
\end{abstract}

\date{\today}

\maketitle
\section{Introduction}
One of the most challenging goals of quantum information theory is to understand what can be accomplished by spatially separated parties, each performing local operations (LO) on quantum subsystems and exchanging classical communication (CC) amongst themselves, a process known as LOCC \cite{Walgate,BennettPurifyTele,BennettConcentrate,CiracEnt,Nielsen,Anselmi,HillerySecret,DBerry,ourNLU,Chefles}. Unfortunately, this class of operations is extremely difficult to analyze, and while much progress has been made, often through the study of specific tasks, we still lack a deep intuition about its inner workings. Separable operations (SEP) \cite{Rains}, of which LOCC is a strict subset \cite{Bennett9}, has a much simpler mathematical description, but while its study has provided powerful lessons about LOCC, no simple picture has emerged that would allow us to understand the difference between these two important classes of quantum operations in the most general terms.

In this paper, we give a conceptually simple picture that distinguishes between LOCC and SEP, and then we show that it provides a very strong separation between these two classes of quantum operations. We will prove a necessary condition for LOCC and then show that separable operations violates this condition by an arbitrarily large amount, an amount constrained only by the size of the system as measured by the number of parties involved. In previous work showing a gap between LOCC and SEP, studies have been made of specific operational tasks that can be accomplished by SEP but cannot be closely approximated by LOCC \cite{FortescueLo,ChitCuiLoPRA,ChitCuiLoPRL,WinterLeung,ChildsLeung}. The gap we show is of a different sort, one which does not directly address the important question of  how closely a given separable operation can be approximated by LOCC. In contrast, our necessary condition is of an abstract, geometrical nature, which provides a more general (and one may hope, ultimately deeper) understanding of the difference between SEP and LOCC.

Let us begin by recalling what LOCC involves. We may assume the parties have agreed in advance upon a protocol that they will follow. One of the parties, say party $1$ whose system is described by states in Hilbert space $\HC_1$, starts by locally performing a generalized measurement \cite{Kraus} with outcomes corresponding to Kraus operators $K_{i_1}$. That party broadcasts her outcome $i_1$ to the other parties, who according to the agreed upon protocol, all know which of them (call this party $2$) is to measure next. Party $2$ then performs a measurement with outcome $i_2$, described by $K_{i_2}^{(i_1)}$ acting on $\HC_2$ and conditioned on Alice's outcome $i_1$, after which he broadcasts his outcome $i_2$ to all the others. The next party to measure will be $\alpha$ (which could be party $1$ again), performing $K^{(i_1,i_2)}_{i_3}$, and they may continue in this way for an arbitrary (but we assume here, finite) number of rounds. From the fact that the probabilities of outcomes obtained at each stage must always sum to unity, one has that for each and every $n$,
\begin{eqnarray}\label{eqn2001}
	I_\alpha = \sum_{i_n}K^{(\SC^\alpha_n)\dagger}_{i_n}K^{(\SC^\alpha_n)}_{i_n},
\end{eqnarray}
where $I_\alpha$ is the identity operator on $\HC_\alpha$, and $\SC^\alpha_n$ is a collection of indices $\SC^\alpha_n=\{i_1,i_2,\cdots,i_{n\!-\!1};\beta\}$ indicating all outcomes obtained in earlier measurements. The last index in the collection, $\beta$, indicates which party performed this measurement, and when $\beta\ne\alpha$ we define $K_{i_n}^{(\SC^\alpha_n)}=I_\alpha$, reflecting the fact that party $\alpha$ does nothing when party $\beta$ is measuring. When this is the case, the sum on the right has only this single term $I_\alpha$, and \eqref{eqn2001} becomes trivial. 

Given any LOCC protocol, we can represent it as a tree, each local measurement appearing as a branching to a set of nodes, with each of these nodes representing one outcome of that measurement. We will label each node by a positive operator $\KC_{i_n}^{(\SC^\alpha_n)}$ obtained as follows: starting from the ordered product of all Kraus operators implemented by the party whose outcome is represented by the given node, $K_{i_n}^{(\SC^\alpha_n)}K_{i_{n\!-\!1}}^{(\SC^\alpha_{n\!-\!1})}\ldots K_{i_2}^{(\SC^\alpha_2)}K_{i_1}^{(\SC^\alpha_1)}$, multiply this by its Hermitian conjugate to obtain the desired positive operator,\begin{align}\label{eqn2002}
\KC_{i_n}^{(\SC^\alpha_n)}=K_{i_1}^{(\SC^\alpha_1)\dag}K_{i_2}^{(\SC^\alpha_2)\dag}\ldots K_{i_{n\!-\!1}}^{(\SC^\alpha_{n\!-\!1})\dag}K_{i_n}^{(\SC^\alpha_n)\dag}K_{i_n}^{(\SC^\alpha_n)}K_{i_{n\!-\!1}}^{(\SC^\alpha_{n\!-\!1})}\ldots K_{i_2}^{(\SC^\alpha_2)}K_{i_1}^{(\SC^\alpha_1)}.
\end{align}
Note that in this product of Kraus operators, there is one for each round leading up to this node, but many of these operators will be the identity, as parties other than $\alpha$ will have measured at that round along this branch of the tree. We also note that in the following, we will sometimes use the term ``outcome" (of a measurement) to refer to these operators, $\KC_{i_n}^{(\SC^\alpha_n)}$, which label the node associated with that outcome.

As observed in \cite{mySEPvsLOCC,myMany}, \eqref{eqn2001} and \eqref{eqn2002} tell us that
\begin{align}\label{eqn2003}
\sum_{i_n}\KC_{i_n}^{(\SC^\alpha_n)}=\KC_{i_{n\!-\!1}}^{(\SC^\alpha_{n\!-\!1})}.
\end{align}
Using this result, we provided a method of constructing an LOCC protocol from an arbitrary separable operation whenever such a (finite-round) protocol exists \cite{mySEPvsLOCC,myMany}.

For clarity and use in the remaining discussion, the following is how we identify the SEP implemented by a given LOCC protocol. When $\{\SC_n^\alpha,i_n\}$ denotes a leaf of the tree, it identifies a final outcome of the protocol. The operator $\hat\KC_j^{(\alpha)}:=\KC_{i_n}^{\SC_n^\alpha}/\Tr
{\KC_{i_n}^{\SC_n^\alpha}}$, with $\SC^\alpha_n=\{i_1,i_2,\cdots,i_{n\!-\!1};\alpha\}$, is then defined to be party $\alpha$'s part of the $j$th outcome in the SEP implemented by this LOCC protocol. The closest $\beta$-node that is an ancestor to this leaf (ancestors are closer to the root, descendants are further) identifies the operator implemented by party $\beta$ for this final outcome of the LOCC, and is therefore (proportional to) $\hat\KC_j^{(\beta)}$ of the SEP. By doing this for each party, we may determine the product operator $\hat\KC_j=\hat\KC_j^{(1)}\otimes\ldots\otimes\hat\KC_j^{(P)}$ associated with each leaf node. The SEP implemented by this LOCC protocol is then defined by the collection of distinct operators $\{\hat\KC_j\}_{j=1}^N$.\footnote{See Theorem~2 and the accompanying discussion in \cite{myMany} for an explanation of why all our results hold equally well when one's interest is in the Kraus operators implemented by the protocol, rather than just the $\hat\KC_j$.} By Theorem~$1$ of \cite{WinterLeung} and the fact that we are restricting our discussion to finite-round protocols, we may, without loss of generality, assume $N$ is finite.

The method of \cite{mySEPvsLOCC,myMany} constructs an LOCC protocol for a SEP by finding intersections of convex cones formed from subsets of the local operators $\hat\KC_j^{(\alpha)}$ for each party $\alpha$. Consideration of the extreme rays\footnote{A ray is a half-line of the form $\{\lambda\hat\KC_j^{(\alpha)}\vert\lambda\ge0\}$, and we will sometimes refer to $\hat\KC_j^{(\alpha)}$ as a `ray', by which we will mean that this operator generates the ray through multiplication by non-negative scalars, $\lambda$. An extreme ray of a convex cone is a ray that lies in the cone but cannot be written as a positive linear combination of other rays in that cone.} of convex cones generated by these operators will lead us to the main result of this paper. The basic idea underlying our result is that too many extreme rays means that one cannot find enough intersections to piece together the full puzzle into a single protocol that incorporates all of the operators defining the given SEP. Before stating our theorem, let us first discuss how we count the $P$ parties. Consider any LOCC protocol involving $\widetilde P$ parties, with the collection of final outcomes corresponding to the set of positive operators, $\{\hat\KC_j\}_{j=1}^N$. If for each $j=1,\ldots,N$ a given party only does an isometry, so that for this party $\hat\KC_j^{(\alpha)}\propto I_\alpha~\forall{j}$, then that party can simply do those isometries at the end of the protocol. Then it is immediate that an LOCC exists for this SEP on $\widetilde P$ parties if and only if one exists for the SEP on $\widetilde P-1$ parties obtained by simply deleting that one party's local operators from the $\hat\KC_j$. Hence, in proving a necessary condition for LOCC, we need only consider SEPs having operators $\hat\KC_j$ such that each of the $P$ parties has at least one local operator $\hat\KC_j^{(\alpha)}$ that is not the identity, and this is what we will do in the remainder of this paper.

Now we state our main result. Let us count the distinct extreme rays in the convex cone generated by the set of local operators, $\{\hat\KC_j^{(\alpha)}\}_{j=1}^N$ for each $\alpha$, and define this number to be $e_\alpha$.
Then, we have the following theorem.
\begin{thm1} \label{thm1}For any finite-round LOCC protocol of $P$ parties implementing a separable operation defined by the $N$ distinct positive product operators $\{\hat\KC_j=\hat\KC_j^{(1)}\otimes\ldots\otimes\hat\KC_j^{(P)}\}_{j=1}^N$, it must be that
\begin{align}\label{eqn101}
\sum_{\alpha=1}^P e_\alpha\le2(N-1),
\end{align}
where $e_\alpha$ is the number of distinct extreme rays in the convex cone generated by operators $\{\hat\KC_j^{(\alpha)}\}_{j=1}^N$, and the sum includes only those parties for which at least one of these local operators is not proportional to the identity. The upper bound in \eqref{eqn101} can be achieved with equality when $N\le2^P$.
\end{thm1}

In the next section we give a proof of this theorem, and then we offer our conclusions. In Appendix~\ref{app3}, we construct separable operations for every $P\ge2$, each one having a unique representation in terms of product Kraus operators, and which satisfy $\sum e_\alpha=PN$, the maximum possible value of this sum, showing that the bound in the theorem is extensively violated by separable operations. The uniqueness of the product representation for each of these separable operations implies that any implementation by LOCC must be in terms of that specific representation's set of Kraus operators, since non-product representations cannot be implemented by LOCC. It therefore follows that in the sense of this theorem, these separable operations are as far from LOCC as possible.

\section{Proof of Theorem~\ref{thm1}}
In this section, we prove our main theorem. The proof will use a series of preliminary results, which will allow us to restrict the LOCC protocols that need to be considered. We begin with the following fairly straightforward result, which is Lemma 4 in \cite{myMany}.
\begin{lem3}\label{lem3}
For any LOCC protocol involving local measurements having two or more proportional outcomes, there is a corresponding LOCC protocol with no two outcomes of any measurement proportional to each other, but which implements the exact same separable operation as the original protocol, including reproducing the same weights for each positive operator $\hat\KC_j$ defining the separable operation.
\end{lem3}
\noindent In addition, if a party performs an isometry (``measurement" with only one outcome) at any stage of an LOCC protocol, they could just as well have absorbed that isometry into their subsequent measurement, omitting the round in which they had implemented the isometry (see the paragraph following Lemma~$1$ in \cite{myMany} for details). Therefore, we can restrict consideration to LOCC protocols for which every round involves a measurement having at least two outcomes.

It is then a simple matter to revise any such LOCC protocol into another for which every measurement has exactly two outcomes, with the latter implementing the exact same separable operation as the former. This can be done with a replacement of each measurement having more than two outcomes by a sequence of measurements having exactly two outcomes each. It is also readily demonstrated that if the original protocol has no two proportional outcomes in any individual measurement, then one can choose each replacement sequence of two-outcome measurements such that none of these measurements has its two outcomes proportional to each other.

These observations tell us that in proving necessary conditions for LOCC, we can (and will do so in the remainder of this paper) restrict consideration to the following special class of LOCC protocols.
\begin{def1}\label{def1}
A ``canonical" LOCC protocol is one where every local measurement has exactly two outcomes, and in each of these measurements, the two outcomes are not proportional to each other. Every such protocol can therefore be represented by a tree with every non-leaf node having exactly two child nodes (these are commonly known as full binary trees). Each node is labeled by a positive operator as shown in \eqref{eqn2002}, and for any given node, the positive operators representing its two child nodes are not proportional to each other. We will refer to such trees as canonical LOCC trees.
\end{def1}
As we will be dealing with full binary trees, the following well-known theorem will be useful.
\begin{thm2}\label{thm2}\cite{FullBinaryTreeThm}
For a full binary tree, the number of leaf nodes exceeds the number of non-leaf nodes by exactly $1$. 
\end{thm2}
\noindent We will also need the following lemma, proved in Appendix~\ref{app1}, concerning the location of nodes in a canonical LOCC tree that may be extreme rays.
\begin{lem1}\label{lem1}
In a canonical LOCC tree, a node $n$ representing an outcome of a measurement by party $\alpha$ cannot be an extreme ray if there is an $\alpha$-node that is a descendant of node $n$.
\end{lem1}
\noindent We are now ready to prove Theorem \ref{thm1}. 

\textit{Proof of Theorem \ref{thm1}.} Consider first the special case where, in a finite canonical LOCC tree, each distinct outcome (each $\hat\KC_j$) appears once and only once as a leaf of the tree. Then, the number of leaves on the tree is equal to the number of outcomes, $N$. Since the protocol begins with all parties having yet to make a measurement, we may label the root of the tree as $I_\alpha$ for some $\alpha$ (it is immaterial for our purposes which $\alpha$ is chosen). Now, $I_\alpha$ is not an extreme ray except when it is the \textit{only} ray in the cone, which can only happen if for that party $\alpha$, $\KC_j^{(\alpha)}=c_jI_\alpha~\forall{j}$. However, in this case we do not include this party in the count of extreme rays (see paragraph preceding Theorem~\ref{thm1}). Therefore, since the root node is not extreme, and since every extreme ray (indeed, every local operator $\hat\KC_j^{(\alpha)}$) in the corresponding SEP is represented by a node in the tree, $\sum_\alpha e_\alpha$ cannot exceed one less than the number of nodes. Since the total number of nodes is $2N-1$ by the full binary tree theorem, we find that $\sum_\alpha e_\alpha\le2(N-1)$ as claimed.

In general, however, there will be repeated outcomes: multiple leafs will correspond to the same outcome of the SEP (the same $\hat\KC_j$). Therefore, we need a way to count extreme rays without counting repetitions of those already counted. We will do this by removing nodes in a way in which the tree remains a full binary tree at every stage of the process, and which leaves at least one instance of each extreme ray. Note that by removing nodes, the remaining tree will no longer correspond to an LOCC protocol, but this is unimportant as our only purpose is to count everything in an appropriate manner. 

Depict the tree with the root at the top and branches extending downward to the right and left. Each non-leaf node is parent to two child nodes, each of which is, in turn, the root of what we may refer to as a child sub-tree of that parent. Notice that every pair of nodes has a closest common ancestor. If that ancestor is not one of the pair, the node that is in the right child sub-tree of this common ancestor is `to the right' of the other, which is necessarily in the left child sub-tree (according to this definition, a node that is an ancestor to another node is neither to the right nor to the left of that other one). For each $j$, there is therefore a right-most $\hat\KC_j$ leaf (a leaf is never ancestor to another leaf), which we choose as the `keeper' $\hat\KC_j$ leaf. In this way we obtain $N$ keeper leafs, where for each $j$ any non-keeper $\hat\KC_j$ leaf is to the left of its respective keeper leaf. All non-keeper leafs will be removed in a way that leaves a full binary tree with $N$ leaf nodes. 

\begin{figure}
\includegraphics[scale=0.6]{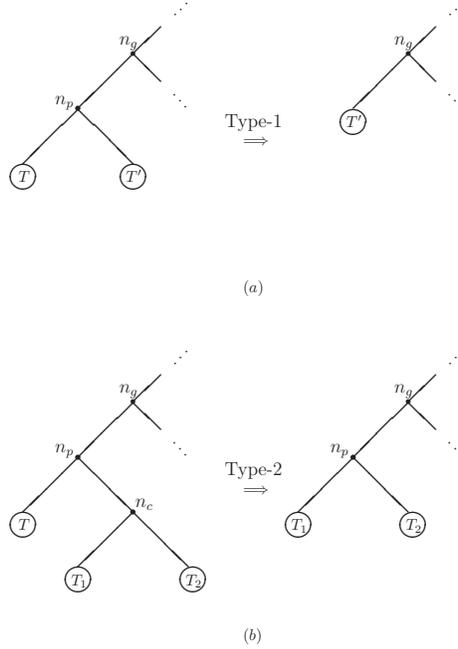}
\caption{\label{fig1}Illustration of the two types of removals used in pruning a canonical LOCC tree. (a) A type-$1$ removal, where the parent $n_p$ is removed along with the maximal keeperless sub-tree $T$. This type of removal is used when there is at least one $\hat\KC_j$ leaf in $T$ whose corresponding keeper is not in $T^\prime$. The root of $T^\prime$ is $n_c$ in the terminology used in the main text; $n_c$ may be the only node in $T^\prime$, in which case it turns out that $n_c$ is itself a keeper leaf.  (b) A type-$2$ removal where $n_c$, which is the root of the sibling sub-tree of $T$, is removed with $T$. This type of removal is used when every leaf in $T$ has its corresponding keeper leaf in either $T_1$ or $T_2$. Under these circumstances, $n_c$ cannot be a leaf.}
\end{figure}

As discussed in the next paragraph, non-keeper leafs will be removed as part of a sub-tree, and this is done in two different ways, which we now describe. Consider a sub-tree $T$ of the full tree, where $T$ has no keeper leafs in it. Denote the parent of $T$ as $n_p$, the other child of this parent as $n_c$. Of the collection of $\hat\KC_j$ leafs in $T$ (with $j$ ranging over all values present in $T$), if at least one of the corresponding keeper $\hat\KC_j$ leafs is not a descendant of $n_p$, then remove $n_p$ along with the entire sub-tree, $T$. The tree is kept as a full binary tree by adding an edge from the parent of $n_p$ to the remaining child node $n_c$. We will refer to this as a `type-$1$' removal. `Type-$2$' removals will be used when every leaf in $T$ has its corresponding keeper leaf as a descendant of $n_p$, and thus of $n_c$, in which case we will remove $n_c$ along with $T$, re-attaching the children of $n_c$ as children, still siblings, of $n_p$, and taking care to preserve the right/left relationship between these children (under these circumstances, it turns out that $n_c$ cannot be a leaf, see Appendix~\ref{app7} for a proof). Again, the tree remains full binary. See Figure~\ref{fig1} for an illustration of these two types of removals.\footnote{\label{foot3}Notice that the tree structure induces a partial order amongst the nodes, having to do with whether or not two nodes are ancestor/descendant of one another. As should be clear from Figure~\ref{fig1}, if a pair of nodes are (are not) ancestor/descendant of one another after a removal, then they were (were not) ancestor/descendant before the removal.} They are chosen to guarantee that there is at least one instance of every extreme ray still present in the fully pruned tree, a fact that will be proven below.

The following is how we will prune the tree. At every stage including the first, consider the left-most non-keeper leaf. If the sibling sub-tree of this non-keeper leaf has no keepers, consider instead the entire sub-tree for which the parent of these sub-trees is the root (that is, consider both sibling sub-trees and their parent as a single sub-tree). Then, if the sibling sub-tree of this larger sub-tree has no keepers combine these two sub-trees with their parent, and consider this larger sub-tree. Continue in this fashion until a keeper leaf is encountered in the sibling sub-tree, having thus found a `maximal' keeperless sub-tree. Remove this entire maximal sub-tree as either type-$1$ or type-$2$, whichever is appropriate. Then, find the left-most non-keeper leaf in the tree that remains, and repeat this process until all non-keeper leafs have been removed.

Having removed all non-keeper leafs, the $N$ keepers are the only leafs remaining in the fully pruned tree, as desired. Furthermore, a non-leaf node is only removed if it is within a sub-tree that has no keeper leaf in it, or if it is that extra non-leaf node that is removed along with one of those sub-trees. This implies that the root of the entire original tree is never removed: the only sub-tree it is within is the full original tree, which obviously has a keeper; and if either child sub-tree of the root has no keeper leaf, then the other sub-tree has in it every keeper corresponding to the non-keepers in that first sub-tree, so the first sub-tree is removed as type-$2$, in which case it is the root of the other sub-tree that is removed as the extra non-leaf, rather than the root of the entire tree.  Hence, the root of the entire original tree is still present as the root of the entire fully pruned tree.

We now prove that one instance of each extreme ray always remains in the fully pruned tree. Suppose, by contradiction, that for some fixed $\alpha$ and each $j\in\JC$ with index set $\JC\subseteq\{1,2,\ldots,N\}$, $\hat\KC_j^{(\alpha)}=:\hat\KC_\ast^{(\alpha)}$ is a (single) extreme ray whose every appearance is removed in our procedure for pruning the tree. This means that no keeper leaf can be $\hat\KC_\ast^{(\alpha)}$, since keeper leafs are never removed. Therefore, each $\hat\KC_j$ keeper leaf with $j\in\JC$ has a $\hat\KC_\ast^{(\alpha)}$ node as its ancestor in the original tree. Recall that each non-keeper is to the left of its respective keeper leaf and consider the right-most appearance of $\hat\KC_\ast^{(\alpha)}$ in the original tree.\footnote{With $\hat\KC_\ast^{(\alpha)}$ assumed to be extreme, no $\hat\KC_\ast^{(\alpha)}$ node can have an $\alpha$-node descendant, according to Lemma~\ref{lem1}. Therefore, no one of these nodes can be ancestor to another one, implying that each $\hat\KC_\ast^{(\alpha)}$ node is either to the right or left of every other one, so there is one of them that is furthest to the right.} This right-most appearance will be ancestor to a keeper $\hat\KC_j$ leaf with $j\in\JC$ since otherwise the non-keeper it is ancestor to would be to the right of its respective keeper. Since this $\hat\KC_\ast^{(\alpha)}$ node is ancestor to a keeper, it is not removed as part of an entire sub-tree $T$ ($T$ is only removed if it has no keepers) so it must be removed as the extra non-leaf that is removed along with $T$. For type-$2$ removals, it turns out that extra non-leaf $n_c$ is not extreme (see Appendix~\ref{app7} for a proof) so this $\hat\KC_\ast^{(\alpha)}$ node must be removed as type-$1$, that is, as parent $n_p$ of $T$. For the (generally, partially pruned) tree to which this type-$1$ removal is applied, then according to how we decide which type of removal to use, there exists a $\hat\KC_i$ leaf in $T$, which is a descendant of $n_p$, whose corresponding keeper is not a descendant of $n_p$. Since ancestral relationships are not altered during pruning (for those nodes that remain in the tree),$^{\ref{foot3}}$ this means that in the original tree, this non-keeper $\hat\KC_i$ leaf in $T$ is a descendant of $n_p$ and its corresponding keeper is not. Thus, the keeper $\hat\KC_i$ leaf is to the right of $n_p$ in the original tree, since that keeper is to the right of a descendant of $n_p$ (that non-keeper) and is not itself a descendant of $n_p$, implying that keeper is in the right child sub-tree of its closest common ancestor with $n_p$. Also in the original tree, this keeper $\hat\KC_i$ leaf is either $\hat\KC_i^{(\alpha)}$ or else has a $\hat\KC_i^{(\alpha)}$ ancestor, and in either case this $\hat\KC_i^{(\alpha)}$ node cannot be a descendant of $n_p$ (or else the keeper $\hat\KC_i$ leaf would be a descendant of $n_p$), so is either ancestor to $n_p$ or is to the right of $n_p$. If it is ancestor to $n_p$, $\hat\KC_i^{(\alpha)}\ne\hat\KC_\ast^{(\alpha)}$, because $\hat\KC_i^{(\alpha)}$ has $n_p$ as an $\alpha$-node descendant, so by Lemma~\ref{lem1} cannot be extreme, which $\hat\KC_\ast^{(\alpha)}$ is, by assumption. If, on the other hand, this $\hat\KC_i^{(\alpha)}$ is to the right of $n_p$, we also have that $\hat\KC_i^{(\alpha)}\ne\hat\KC_\ast^{(\alpha)}$, because $n_p$ is the right-most $\hat\KC_\ast^{(\alpha)}$ node. In either case, the nearest $\alpha$-node to that non-keeper $\hat\KC_i$ leaf in child sub-tree $T$ of $n_p$ is $\hat\KC_i^{(\alpha)}\ne\hat\KC_\ast^{(\alpha)}$, so is not $n_p$, implying $T$ must have an $\alpha$-node in it, which is thus a descendant of $n_p$. Since $n_p$ is $\hat\KC_\ast^{(\alpha)}$, Lemma~\ref{lem1} then tells us that $\hat\KC_\ast^{(\alpha)}$ is not extreme, a contradiction, proving that every extreme ray is present at least once in the fully pruned tree.

Thus, our fully pruned tree, which is a full binary tree, has $N$ leaf nodes and $2N-1$ nodes in all, and every extreme ray is present as one of its nodes. Since the root of the original tree is not extreme and is never removed, there can be no more than $2(N-1)$ extreme rays. It is shown in Appendix~\ref{app2} that this bound can be saturated when $N\le2^P$, which completes the proof.\hspace{\stretch{1}}$\blacksquare$

\section{Conclusions}\label{conc}
We have proved a necessary condition for any finite-round LOCC protocol, stated in Theorem~\ref{thm1}. We have also demonstrated (Appendix~\ref{app3}) that this necessary condition is violated extensively by separable operations, this violation growing without bound as the number of parties increases. Considering an arbitrary SEP, violating our bound becomes a general sufficient condition that the SEP cannot be exactly implemented by LOCC. We note that for the bipartite case, we have an independent argument that provides a better bound whenever $N>4$, that bound being $\sum_\alpha e_\alpha\le3N/2$. This shows that, at least for $P=2$, the bound of Theorem~\ref{thm1} can only be saturated by finite-round LOCC when $N\le2^P$. In addition, we note that our necessary condition for LOCC is demonstrably not sufficient. An example is the SEP presented in \cite{Bennett9} as the first demonstration that SEP and LOCC are inequivalent. In this example, for which $N=9$ and $P=2$, we have that $e_1=7=e_2$, so that $e_1+e_2=14<16=2(N-1)$. The bound in our theorem is satisfied, but this SEP cannot be exactly implemented by LOCC.

We hope that these ideas will open new avenues toward understanding LOCC. One important outstanding question is whether, for a given SEP, there is a relationship between the extent to which the bound in Theorem~\ref{thm1} is violated and how closely the SEP can be approximated by LOCC. It is also of interest to determine whether or not the bound in Theorem~\ref{thm1} can be violated by infinite-round LOCC, an important question that we have as yet been unable to answer.

~

\noindent\textit{Acknowledgments} --- The author would like to thank Li Yu for helpful suggestions and especially Dan Stahlke for his numerous and very helpful comments and suggestions. This work has been supported in part by the National Science Foundation through Grant No. 1205931.

\appendix
\section{Separable channels on $P$ parties with $\sum_\alpha e_\alpha=PN$}\label{app3}
Here, we construct SEPs as sets of positive operators $\{\hat\KC_j\}$ for every $P$ and for which Theorem \ref{thm1} is violated maximally, having $\sum_\alpha e_\alpha=PN$. Define operators $\hat\KC_j=\dyad{\Psi_j}{\Psi_j},~j=1,\ldots,N$, where $\ket{\Psi_j}=(D/N)^{1/2}\ket{\psi_j^{(1)}}\otimes\ldots\otimes\ket{\psi_j^{(P)}}$, $D=d_1d_2\ldots d_P$, $d_\alpha$ is the dimension of Hilbert space $\HC_\alpha$ with parties ordered such that $d_1\le d_2\le\ldots\le d_P$, and the state on party $\alpha$'s subsystem is
\begin{align}\label{eqn501}
\ket{\psi_j^{(\alpha)}}=\frac{1}{\sqrt{d_\alpha}}\sum_{m_\alpha=1}^{d_\alpha}e^{2\pi \textrm{i}jp_\alpha m_\alpha/N}\ket{m_\alpha}.
\end{align}
Here, $p_1=1$ and for $\alpha\ge2$, $p_\alpha=d_1d_2\ldots d_{\alpha-1}$, $\ket{m_\alpha}$ is the standard basis for party $\alpha$, and $N$ is chosen as any prime number exceeding $D$. Since for each positive operator $\hat\KC_j$ the local parts are the rank-$1$ projectors $\hat\KC_j^{(\alpha)}=\ket{\psi_j^{(\alpha)}}\bra{\psi_j^{(\alpha)}}$, each $\hat\KC_j^{(\alpha)}$ is thus an extreme ray of its respective convex cone. This means that $e_\alpha=N~\forall{\alpha}$ and $\sum_\alpha e_\alpha=PN$, an extensive violation of Theorem~\ref{thm1} and the maximal possible value of this sum. 

We need to show that the set $\{\hat\KC_j\}$ satisfies closure, $\sum_j\hat\KC_j=I$. We have,
\begin{align}\label{eqn601}
\sum_{j=1}^N \hat\KC_j&=\frac{1}{N}\sum_{m_1,n_1=1}^{d_1}\ldots\sum_{m_P,n_P=1}^{d_P}\left(\sum_{j=1}^Ne^{2\pi \textrm{i}j\sum_\alpha p_\alpha (m_\alpha-n_\alpha)/N}\right)\dyad{m_1\ldots m_P}{n_1\ldots n_P}\notag\\
&=\sum_{m_1,n_1=1}^{d_1}\ldots\sum_{m_P,n_P=1}^{d_P}\delta\left(\sum_{\alpha=1}^P p_\alpha m_\alpha,\sum_{\alpha=1}^P p_\alpha n_\alpha\right)\dyad{m_1\ldots m_P}{n_1\ldots n_P},
\end{align}
where $\delta(\cdot,\cdot)$ is the Kronecker delta, vanishing unless its two arguments are equal, in which case it is equal to unity. Recall that $p_1=1$ and $p_\alpha=d_1d_2\ldots d_{\alpha-1}$, $\alpha\ge2$. Suppose $m_P\ne n_P$. Then, $p_P|m_P-n_P|\ge p_P=d_1d_2\cdots d_{P-1}>\sum_{\alpha=1}^{P-1} p_\alpha |m_\alpha-n_\alpha|$, since the right-hand side of this inequality is no greater than $d_1-1+d_1(d_2-1)+d_1d_2(d_3-1)+\ldots+d_1d_2\cdots d_{P-2}(d_{P-1}-1)=d_1d_2\cdots d_{P-1}-1$. We conclude that equality of the two arguments in the Kronecker delta in the last line of \eqref{eqn601} requires $m_P=n_P$. Similar arguments, proceeding sequentially with decreasing $\alpha$ starting next from $\alpha=P-1$, shows that $m_\alpha=n_\alpha~\forall{\alpha}$. Thus, the right-hand side of \eqref{eqn601} is equal to $I$, the identity on the full input Hilbert space, as desired.

Finally for each $P$, we show the existence of sets of Kraus operators corresponding to the positive operators given in the previous paragraph, which are the unique product Kraus representation for their associated quantum channel. Define Kraus operators $\hat K_j=\ket{\Phi_j}\bra{\Psi_j}$, with $\ket{\Psi_j}$ defined above \eqref{eqn501}. Define normalized states $\ket{\Phi_j}=\ket{\phi_j^{(1)}}\otimes\ket{\Phi_j^\prime}$ with $\{\ket{\Phi_j^\prime}\}_{j=1}^N$ a set of linearly independent product states on the $P-1$ parties excluding party $1$ (this requires that at least one output dimension of those last $P-1$ parties exceeds its input, in order that the overall output dimension is not less than $N$, the number of these independent states). Then, since no two of the $\ket{\psi_j^{(1)}}$ are proportional to each other, the conditions of Corollary $1$ of \cite{mySumOfProd} are met for this set of Kraus operators (with a bipartite split between party $1$ and all the rest), which implies that no linear combination of the $\hat K_j$ is a product operator (apart from the $\hat K_j$'s themselves). This, in turn, implies that the set $\{\hat K_j\}_{j=1}^N$ is the unique product representation for the given channel. Therefore, there is no other Kraus representation that could possibly be LOCC, and these channels are as far from LOCC as possible, in the sense of Theorem~\ref{thm1} of the main text, as claimed.

\section{Proof of Lemma~\ref{lem1}}\label{app1}
If $\alpha$-node $n$ has another $\alpha$-node that is its descendant, then party $\alpha$ made a measurement after the measurement that produced node $n$. Then, by \eqref{eqn2003}, the positive operator labeling node $n$ is a sum of those positive operators labeling the descendant $\alpha$-nodes produced by the later measurement. In turn, some of those descendants may themselves have other descendant $\alpha$-nodes produced by a subsequent measurement, so are sums of those operators corresponding to the latter nodes. Eventually if we continue toward the leaves, we end up with nodes that are labeled by the $\hat\KC_j^{(\alpha)}$ defining the separable operation, so that node $n$ is seen to be labeled by a sum of those $\hat\KC_j^{(\alpha)}$. This means node $n$ cannot be an extreme ray in the convex hull of the $\hat\KC_j^{(\alpha)}$ unless all the $\hat\KC_j^{(\alpha)}$ entering this sum are proportional to each other, which is impossible in the canonical LOCC trees we are considering.\hspace{\stretch{1}}$\blacksquare$

\section{Proof that $n_c$ is a non-leaf that is not extreme for all type-$2$ removals}\label{app7}

\begin{figure}
\includegraphics[scale=0.6]{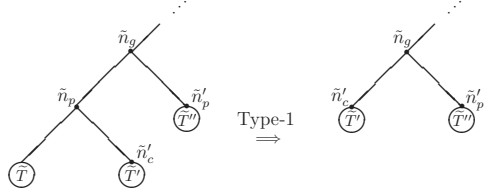}
\caption{\label{fig2}A type-$1$ removal creates a new sibling pair, $\tilde n_c^\prime$ and $\tilde n_p^\prime$, but this pair will never be involved in a subsequent type-$2$ removal, showing that type-$2$ removals always involve siblings that were siblings in the original tree.}
\end{figure}

For type-$2$ removals (of sub-tree $T$ with present sibling $n_c$), we need to show that $n_c$ is a non-leaf that is not extreme. Our argument will utilize the fact that $n_c$ and the root of $T$ were siblings in the original tree, so we need to be sure this is always the case, even following earlier removals. The only way siblings change during the pruning process is via type-$1$ removals, since type-$2$ removals do not change sibling relationships. Suppose there was a type-$1$ removal of $\widetilde T$, child sub-tree of $\tilde n_p$, where the other child sub-tree of $\tilde n_p$ was $\widetilde T^\prime$ and the sibling of $\tilde n_p$ before this removal was $\tilde n_p^\prime$. This removal only changes sibling relationships by changing the sibling of $\tilde n_p^\prime$ from $\tilde n_p$ to the root of $\widetilde T^\prime$, which we denote as $\tilde n_c^\prime$, see Figure~\ref{fig2}. 

We now argue that there will never be a type-$2$ removal involving nodes at the positions of this newly created sibling pair, even if their identities change further through subsequent pruning. Since $\widetilde T^\prime$ presently has a keeper leaf (because otherwise the removed $\widetilde T$ was not a maximal keeperless sub-tree, which are the only ones we remove), it will always have a keeper no matter how much further pruning occurs, since keepers are never removed, so a type-$2$ removal involving these nodes can only occur if the new sibling sub-tree of $\widetilde T^\prime$ (call this sub-tree $\widetilde T^{\prime\prime}$) is keeperless, which means it was keeperless to begin with, even before the previous type-$1$ removal that changed the sibling pair. This means that $\widetilde T^{\prime\prime}$ is to the right of $\widetilde T^\prime$, because otherwise $\widetilde T^{\prime\prime}$ would have been a maximal keeperless sub-tree, removed before that previous type-$1$ removal, since the pruning proceeds from left to right. However, if it is to the right, then the keeper leafs corresponding to the non-keepers in $\widetilde T^{\prime\prime}$ are further to the right and therefore not in $\widetilde T^\prime$, so removal of $\widetilde T^{\prime\prime}$ will be via a type-$1$ removal, not type-$2$, and according to the above discussion, this is true no matter what pruning takes place between the previous type-$1$ removal and this one. Therefore, all type-$2$ removals involve siblings that were siblings in the original tree.

Consider a type-$2$ removal of $T$, a child sub-tree of $n_p$, with $n_p$'s other child $n_c$ removed along with $T$. We can now prove that $n_c$ is not extreme. From the discussion above, we know that the root of $T$ was sibling to $n_c$ in the original (canonical) tree, so these two are not proportional and are both $\alpha$-nodes for the same $\alpha$.
Assume, by contradiction, $n_c$ is extreme. Then by Lemma~\ref{lem1}, $n_c$ is an $\alpha$-node with no $\alpha$-node descendants, 
 so $n_c$ must be proportional to $\hat\KC_j^{(\alpha)}$ for every $j$ such that $\hat\KC_j$ is one of the keeper leafs that are descendants of $n_c$. 
Since every leaf in $T$ has its corresponding keeper leaf as a descendant of $n_c$, 
the closest $\alpha$-node to each and every leaf in $T$ must also be proportional to this same $\hat\KC_j^{(\alpha)}$ (since the root of $T$ is an $\alpha$-node, there is at least one such node in $T$). Therefore by \eqref{eqn2003} of the main text, every $\alpha$-node in $T$, including its root node, is proportional to this same $\hat\KC_j^{(\alpha)}$. In other words, the root of $T$ is proportional to $n_c$, a contradiction, proving that $n_c$ is not extreme. Repeating the exact same argument starting from the assumption $n_c$ is a leaf leads to the same contradiction, thus proving $n_c$ is a non-leaf, and we are done.

\section{Saturating the bound in Theorem~\ref{thm1}}\label{app2}
We will now show that any LOCC protocol satisfying the following two conditions saturates the bound in Theorem~\ref{thm1}, $\sum_\alpha e_\alpha=2(N-1)$:
\begin{enumerate}
  \item \label{itema1}Each party measures once and only once with the same ordering of the parties no matter which outcomes were obtained by the preceding parties. In addition, each of these measurements has exactly two outcomes.

Along any branch of the associated LOCC tree, party $1$ starts with a two-outcome measurement, followed by a two-outcome measurement by party $2$, which is followed by a two-outcome measurement by party $3$, and so on until each party has measured once, party $P$ always making the final measurement. For the entire protocol, party $\alpha$ has $2^{\alpha-1}$ different measurements, which measurement that party makes being determined by the outcomes of all previous parties' measurements. Party $\alpha$ has a total of $2\cdot2^{\alpha-1}=2^\alpha$ measurement outcomes.
  \item Each of the $2^\alpha$ outcomes for party $\alpha$ is a distinct extreme ray in the cone generated by the collection of these outcomes. Note that for a protocol satisfying the preceding condition \ref{itema1}, many of the $\hat\KC_j^{(\alpha)}$ (fixed $\alpha$ but different $j$) will be equal to each other, which explains why party $\alpha\ne P$ has only $2^\alpha$ extreme rays, rather than the maximum possible number, $N=2^P$ .
\end{enumerate}

Here is one specific example that does the trick. For each of party $\alpha$'s measurements, indexed by $m=1,\ldots,2^{\alpha-1}$, let one Kraus operator be a projector onto (normalized) pure state $\ket{\xi_m^{(\alpha)}}$, with the other outcome $I_\alpha-\ket{\xi_m^{(\alpha)}}\bra{\xi_m^{(\alpha)}}$. As these are both projectors, the $\hat\KC_j^{(\alpha)}$ are equal to these Kraus operators, 
\begin{align}\label{eqn701}
\hat\KC_{2m\!-\!1}^{(\alpha)}&=\ket{\xi_m^{(\alpha)}}\bra{\xi_m^{(\alpha)}}\notag\\
\hat\KC_{2m}^{(\alpha)}&=I_\alpha-\ket{\xi_m^{(\alpha)}}\bra{\xi_m^{(\alpha)}}.
\end{align}
Choose the set of pure states $\ket{\xi_m^{(\alpha)}}$ in each $\HC_\alpha$ such that no two are the same and no two are orthogonal to each other. Then, the following argument shows that each and every $\hat\KC_j^{(\alpha)}$ is an extreme ray in the convex cone generated by the collection of all of them (many are repeated, as explained above): The pure state projectors $\hat\KC_{2m\!-\!1}^{(\alpha)}$ are each an extreme ray in the cone of the full set of positive operators acting on $\HC_\alpha$, so are necessarily also extreme in the cone of the collection of $\hat\KC_j^{(\alpha)}$. If $\HC_\alpha$ is two-dimensional, then $\hat\KC_{2m}^{(\alpha)}$ is also extreme for each $m$, by the same argument. Otherwise, $\hat\KC_{2m}^{(\alpha)}$ has rank exceeding unity so is not extreme in the cone of positive operators, but is nonetheless on the boundary of that set, since $\hat\KC_{2m}^{(\alpha)}\ket{\xi_m^{(\alpha)}}=0$. Note also that $p_j:=\bra{\xi_m^{(\alpha)}}\hat\KC_j^{(\alpha)}\ket{\xi_m^{(\alpha)}}>0$ for every $j\ne2m$. Suppose $\hat\KC_{2m}^{(\alpha)}=\sum_{j\ne2m} c_j\hat\KC_j$, with $c_j\ge0$. Taking the diagonal element, $\bra{\xi_m^{(\alpha)}}\cdots\ket{\xi_m^{(\alpha)}}$, of this equation leads to $0=\sum_{j\ne2m}c_jp_j$, or $c_j=0~\forall{j\ne2m}$, a contradiction since $\hat\KC_{2m}^{(\alpha)}\ne0$. Therefore, $\hat\KC_{2m}^{(\alpha)}$ cannot be written as a positive linear combination of all the others, so is an extreme ray in the cone of their collection. We conclude that each and every $\KC_j^{(\alpha)}$ is an extreme ray, $e_\alpha=2^\alpha~\forall{\alpha}$, and $\sum_\alpha e_\alpha=2+2^2+\ldots+2^P=2(2^P-1)=2(N-1)$, saturating the bound.

If the last party omits his measurement for one of the outcomes of the next-to-last party, this removes a pair of leaf nodes, replacing it with a single leaf. Therefore, this reduces $N$ by unity to $2^P-1$. The number of extreme rays is reduced by two, since the new leaf was counted as extreme before the pair of leafs was removed, so the bound is still saturated. By continually omitting single measurements by the last party to measure along a given branch, $N$ is reduced by unity for each omission, and the number of extreme rays is reduced by two. In this way, we can obtain examples saturating the bound for any $N$ satisfying $P+1\le N\le2^P$ [$N$ cannot be less than $P+1$, since all $P$ parties perform a non-trivial measurement at least once in the full protocol; alternatively, one may note that according to the way we count parties, $e_\alpha\ge2~\forall{\alpha}$, so $2P\le\sum e_\alpha\le2(N-1)$, which also shows $N\ge P+1$]. This completes the proof of Theorem~\ref{thm1}.


\end{document}